# iMOLSDOCK : induced-fit docking using mutually orthogonal Latin squares (MOLS)

*D. Sam Paul and N. Gautham*[*]

* Corresponding Author – Dr. N. Gautham, n_gautham@hotmail.com

Centre of Advanced Study in Crystallography and Biophysics, University of Madras, Chennai 600025. India

**KEYWORDS**

Molecular docking, Induced-fit docking, peptide-protein docking, mutually orthogonal Latin squares, receptor flexibility, side-chain flexibility, docking tool, MOLSDOCK

**ABSTRACT**

We have earlier reported the MOLSDOCK technique to perform rigid receptor/flexible ligand docking. The method uses the MOLS method, developed in our laboratory. In this paper we report iMOLSDOCK, the 'flexible receptor' extension we have carried out to the algorithm MOLSDOCK. iMOLSDOCK uses mutually orthogonal Latin squares (MOLS) to sample the conformation and the docking pose of the ligand and also the flexible residues of the receptor protein. The method then uses a variant of the mean field technique to analyze the sample to arrive at the optimum. We have benchmarked and validated iMOLSDOCK with a dataset of 44 peptide-protein complexes with peptides. We have also compared iMOLSDOCK with other flexible receptor docking tools GOLD v5.2.1 and AutoDock Vina. The results obtained show that the method works better than these two algorithms, though it consumes more computer time.

## Introduction

Designing organic small molecules that ‘fit’ into biological receptors is of great interest in drug discovery research. Computational molecular docking is frequently used to imitate the behaviour of a ligand (small peptide or small molecule) in the active pocket of receptor proteins. The earliest method used for automated ligand docking was the DOCK program [1] which held the conformation of the ligand and the receptor rigid but varied the orientation degrees of freedom[2]. After that, docking algorithms which take into account conformational degrees of the ligand as well of the receptor have been developed [3–5].

We had developed the MOLSDOCK technique[6,7] to perform rigid receptor/flexible ligand docking. The method uses the MOLS technique, earlier developed in our laboratory[8]. Using mutually orthogonal Latin squares[9], the algorithm systematically identifies a small but completely representative sample of the enormous multidimensional search space. The energy values are calculated at each of the sampled points. These energy values are then analysed using a variant of the mean field method[10] to simultaneously obtain the optimal conformation, position and orientation of a small molecule ligand on a rigid protein receptor.

However, in the living cell, receptor proteins are not rigid, but may flex and move to accommodate the ligand[11–13]. Significant differences in side-chain conformations occur in at least 60% of the proteins between the *apo* and *holo* forms upon ligand binding, though main-chain conformations are largely preserved [14]. In ~85% of the cases, side-chain changes were observed in only three residues or less in the binding pocket [15]. Thus it may be useful to allow for receptor flexibility, especially side-chain flexibility, as much as for the possible conformational changes in the ligand.

Some of the docking programs that allow receptor flexibility are AutoDock Vina [16], GOLD [17], GLIDE [18] and RosettaLigand [19]. AutoDock Vina [16] is a new generation docking in which a succession of steps consisting of a mutation and a local optimization are taken, with each step being accepted according to the Metropolis criterion. Receptor side-chains may be chosen as flexible. GOLD [17] is a genetic algorithm-based docking tool which allows for partial protein flexibility. GLIDE (grid-based ligand docking with energetics) uses exhaustive systematic search approach, with approximations and truncations, to find the optimal ligand pose. The Induced Fit Docking protocol in GLIDE docking tool [18] begins by docking the active ligand using the GLIDE program. Prime, a protein structure prediction technique, is then used to accommodate the ligand by reorienting nearby side-chains in the active site of the protein. The initial implementation of RosettaLigand[19] was based on Rosetta [20], which allowed the rigid body position and orientation of the ligand and the side-chain conformations in the active site of the protein to be optimized simultaneously using a Monte Carlo minimization procedure. Later RosettaLigand included backbone flexibility [21] with full side-chain flexibility, along with simultaneous treatment of the ligand and the protein. The latest release of RosettaLigand has included simultaneous docking of multiple flexible ligands, accounting for receptor flexibility [22].

In this paper we report the ‘flexible receptor’ extension we have carried out to the MOLSDOCK algorithm. We call the upgraded version of MOLSDOCK with receptor flexibility as iMOLSDOCK. The new docking method iMOLSDOCK, uses ‘induced-fit’ side-chain fluctuations in the selected flexible residues of the receptor protein, to account for receptor flexibility. To benchmark iMOLSDOCK, we have applied this method to 44 test cases (protein-peptide complexes) selected from the Protein Data Bank (PDB). We have compared the performance of the present algorithm with popular flexible receptor docking tools : AutoDock Vina[16] and GOLD v5.2.1[17]. We present the results here.

**Materials and Methods**

MOLSDOCK[6,7,23] is a 'rigid receptor/flexible ligand' docking method developed using the MOLS method. The MOLS method has been presented in detail elsewhere[8,24–26]. Here, we give a brief description for completeness. For simplicity, we will first consider the method as applied to the prediction of the minimum energy structure of a peptide. The conformation space of a peptide may be defined as the set of all possible combinations of all values of all its variable torsion angles. Thus, if there are '*n*' such torsion angles, each which can take up '*m*' different values, there are $m^n$ different combinations of these values. In other words, there are $m^n$ different conformations for the peptide. For each conformation a potential energy may be calculated, and therefore the multi-dimensional conformational space can also be considered to be potential energy surface of the peptide. The task then in peptide structure prediction is to locate the minimum value on this surface, i.e. the minimum energy conformation. A brute force search will lead to combinatorial explosion and, for peptides longer than about 15 residues, is beyond the capabilities of most current-day computers. Here, we use mutually orthogonal Latin squares[9] to systematically choose a set of $m^2$ points (or conformations) from the conformational space of the peptide. The potential energy for each conformation is calculated. The $m^2$ energy values are then analyzed using a variant of the mean field technique[10,27] to arrive at the conformation with the lowest energy. We note here that there are a very large number of ways in which we can choose the $m^2$ points[28]. Each choice can lead to either the same, or to a different low energy structure. Previous experience[6–8,26,29] has shown us that generating about 1,500 structures is sufficient to cover the entire conformational space of a peptide or any other small molecule.  The conformations so generated are clustered together to identify the few lowest energy structures that are possible for the peptide.

The method was later extended to address the docking problem[6,7]. The peptide or organic molecule is considered the flexible ligand, binding to a rigid receptor protein. In addition to the variable torsion angles of the ligand, we included, in the search space, the variables for the position and orientation of the ligand with respect to the protein.  The energy function is also modified to include the interaction energy between protein and ligand, along with the intramolecular energy of the ligand.  The method, dubbed MOLSDOCK, is suitable for all organic small molecules, including peptides, as ligands.

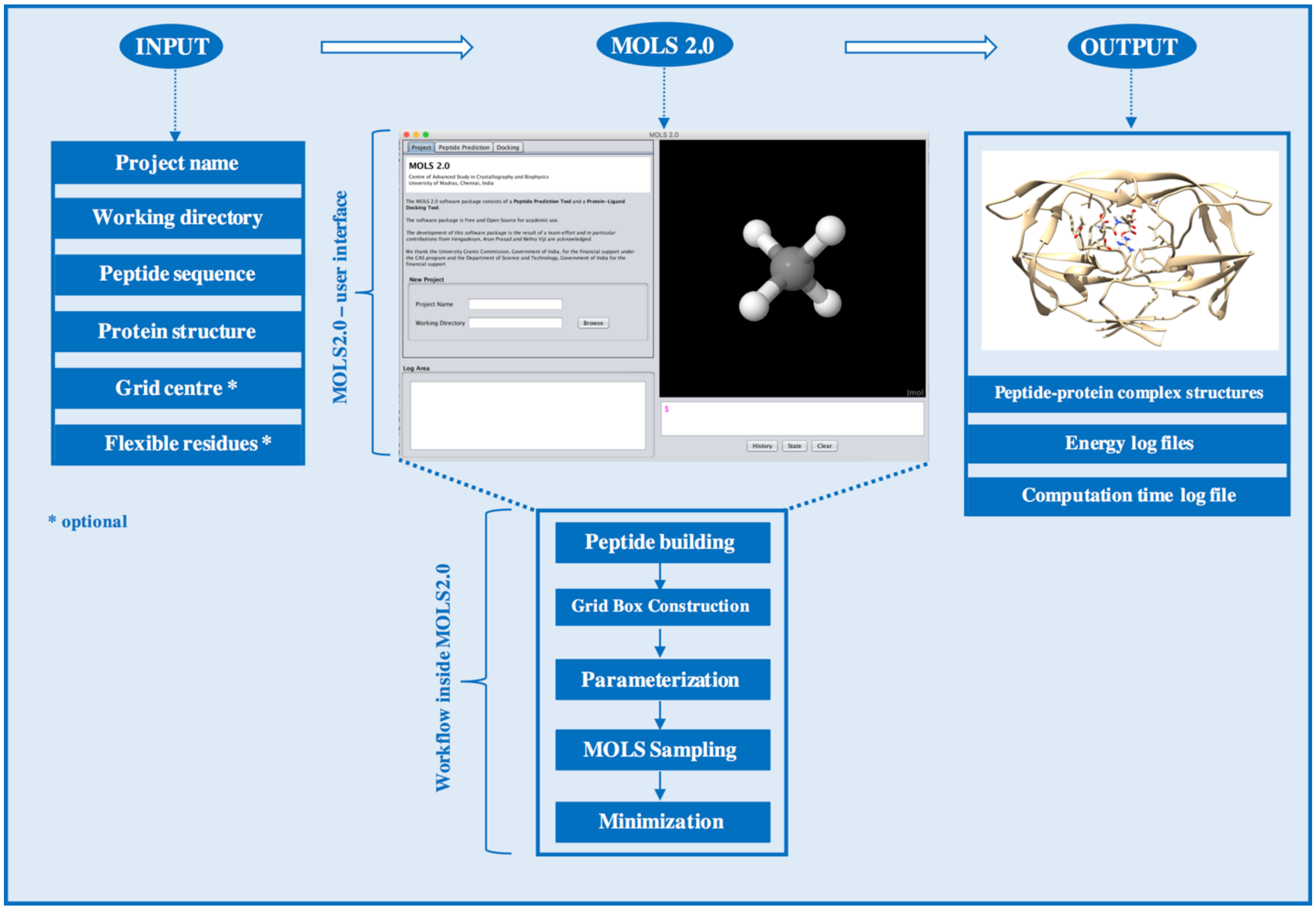

**Figure 1.** The workflow of iMOLSDOCK docking protocol using the MOLS 2.0 interface. If the Grid centre and flexible residues are not known, options are available in MOLS 2.0 to find them automatically.

Now, we have upgraded MOLSDOCK to iMOLSDOCK by including receptor flexibility. Figure 1 shows the work flow of the iMOLSDOCK docking protocol using the MOLS 2.0 interface [30]. The MOLS technique is used to systematically sample not only the potential energy surface of the peptide ligand, and not only the 'docking space' which involves peptide's rotation and translation, but also the conformational space of the side-chains of the flexible residues of the receptor protein which interact with the incoming ligand. (The backbone atoms of the protein are held fixed.)

Two major changes have been made while upgrading MOLSDOCK to iMOLSDOCK. Firstly, the search space is expanded to include the flexible residues of the receptor protein. Secondly, intra-protein energy, which assesses the receptor protein's conformational energy, is added to the scoring function. The conformation of the peptide ligand is specified by '*m*' torsion angles ($\theta_r$, r = 1, *m*), and six parameters describe the peptide's pose, i.e., three for the position and three for the orientation in the receptor binding site. If there are '*p*' torsion angles that describe the flexible residues in the receptor binding site, we have a total of *m+6+p* dimensions in the search space ($\theta_r$, r = 1, *m+6+p*). If each dimension is sampled at '*n'* intervals, then the volume of the search space is $(n)^{m+6+p}$. But the MOLS technique calculates the values of the scoring function only at $(n)^2$ points in this space, and analyses them using a variant of the mean field technique[10,27], to simultaneously identify the optimum conformation of the peptide, its pose, and also the conformation of the side-chains of the protein flexible residues. The search space is defined on a discrete grid of '*m+6+p'* dimensions, and the method identifies an optimum point on the grid. However the actual optimum may lie close to but not actually on the grid. Therefore the final step is to perform a gradient minimization to find the nearest off-grid optimum.

For the peptide conformation, all its variable torsion angles, main chain and side chain, are sampled from 0° to 360° at intervals of 10°. The orientation of the ligand in the receptor site is specified by three angles, two of which specify the position of a rotation axis and the remaining one is the angle of rotation about this axis, as described by Arun Prasad and Gautham (2008). The translational position of the ligand is sampled at intervals of 0.14 Å along three dimensions inside a cubic box of 5 Å centred at the centre of the receptor binding site. This site is either identified from the crystal structure, if known, or may be found using the Fpocket 2.0 algorithm[30,31].

Ligand binding may induce large structural changes in the receptor protein, such as the movement of loops[32,33] or even large domains[34,35]. In most cases, changes in backbone structure are negligible and only side-chain reorientation occurs upon ligand binding[36,37]. Najmanovich et al. (2000) show that backbone displacements on ligand binding is of minor importance when compared to side-chain flexibility. In iMOLSDOCK, we concentrated on the residues lining the binding site and incorporated side-chain flexibility, where side-chains obstructing the binding site are moved to accommodate tight binding of the ligand in the binding site. Observations by Heringa and Argos [38] show that ligand binding induces nonrotamericity in the preferred side-chain conformations. Zavodszky and Kuhn (2005) show that side-chain conformational changes upon ligand binding are typically not changes between rotamers, but mostly involve small (<15°) changes in side-chain dihedral angles relative to the original (typically rotameric) position. Comparison between the ligand-free and ligand-bound X-ray structures show that the side-chain dihedral angles difference is 45° or less in most (~90%) of the cases and is 15° or less in ~80% of the cases [14]. In iMOLSDOCK we have avoided rotameric side-chain exploration and allowed the side-chain torsion angles of the flexible residues to fluctuate within the range 0° and 30° (i.e. -15° to +15° from their position in the crystal structure). Prediction of the ligand docking pose and the receptor flexibility occur simultaneously, resulting in an 'induced-fit' docking in iMOLSDOCK. If the flexible residues that line the active site of the receptor protein are known, they may be specified explicitly. Else residues that are within 4.0 Å from each atom of the peptide ligand are automatically selected. In the test cases (Table 1), we have allowed a maximum of 50 flexible residues in the receptor.

The scoring function in iMOLSDOCK is the sum of the intra-molecular ligand energy, the intra-molecular protein energy and the inter-molecular interaction energy between the ligand and the receptor. Both the intra-molecular protein energy and intra-molecular peptide ligand energy are calculated using the non-bonded terms of the AMBER94 force field[39]. The intermolecular interaction energy between the protein and the ligand is calculated using the PLP scoring function[40]. The total potential energy is the weighted sum of intra-molecular peptide energy, intra-molecular protein energy and the intermolecular protein-peptide interaction energy. The optimum weights were chosen to yield maximum positive correlation between the energy and the root mean square deviation (RMSD) of the resulting docked structure with respect to the 'native' crystal structure. The AMBER force field is reported in units of kcal/mol[39], whereas the PLP force field is reported in dimensionless units[40]. Therefore the total potential energy is reported in dimensionless units.

Earlier studies using MOLS sampling for peptide prediction suggest that a MOLS sample of 1,500 structures is exhaustive[6–8,26,29]. For iMOLSDOCK, we did an analysis to find the number of structures sufficient for an exhaustive sampling. We generated 1,500 structures for all the test cases (Table 1) and then clustered the structures with K-means clustering algorithm using MMTSB toolset [41]. During the clustering procedure, the conformations were considered as different, when the $C_{\alpha}$ atom superposition RMSD between them was greater than 2.0 Å. For the tripeptide cases, 68% of the clusters had at least one representative within the first 100 structures generated. In the case of tetrapeptides and pentapeptides all the clusters had at least one representative structure within the first 100 structures. In the case of hexapeptide, heptapeptide and octapeptide test cases 98%, 95% and 76% of all the clusters had at least one model within the first 100 structures. On an average, for all the test cases, 84% of the clusters had at least one model within the first 100 structures. For each test case, we calculated the occurrence of unique clusters for every 100 models from 1 to 1,500 (Supplementary Figure 1). After the first 1,300 structures, no significant unique clusters occurred. Therefore generating 1,500 structures using MOLS sampling is sufficient to identify all the low energy docked conformations of the ligand in the protein receptor site.

All docking simulations were run under the Linux operating system. The tests we present here were run on protein-peptide complex structures selected from the Protein Data Bank (PDB). We selected entries for which the X-ray structures of both the complex and the apo (unbound) crystal structures were known with a resolution better than 2.40 Å (Table 1). To test successful peptide docking in iMOLSDOCK, it is necessary to dock the peptides using a predefined binding site before blind docking [42]. iMOLSDOCK was applied to *apo* structures with predefined binding sites.

| PDB ID (holo) | Resolution (Å) | PDB ID (apo) | Resolution (Å) | Peptide sequence | No. of torsion angles | Protein |
|---|---|---|---|---|---|---|
| **Tri-peptide complexes** | | | | | | |
| 1A30 | 2.00 | 1G6L | 1.90 | EDL | 13 | HIV-1 protease |
| 8GCH | 1.60 | 4CHA | 1.68 | GAW | 8 | Complex of alpha-chymotrypsin |
| 1B05 | 2.00 | 1RKM | 2.40 | KCK | 15 | Oligopeptide binding protein (oppA) |
| 1B32 | 1.75 | 1RKM | 2.40 | KMK | 17 | Oligopeptide binding protein (oppA) |
| 1B3F | 1.80 | 1RKM | 2.40 | KHK | 16 | Oligopeptide binding protein (oppA) |
| 1B3G | 2.00 | 1RKM | 2.40 | KIK | 16 | Oligopeptide binding protein (oppA) |
| 1B3L | 2.00 | 1RKM | 2.40 | KGK | 14 | Oligopeptide binding protein (oppA) |
| 1B46 | 1.80 | 1RKM | 2.40 | KPK | 13 | Oligopeptide binding protein (oppA) |
| 1B4Z | 1.75 | 1RKM | 2.40 | KDK | 16 | Oligopeptide binding protein (oppA) |
| 1B51 | 1.80 | 1RKM | 2.40 | KSK | 15 | Oligopeptide binding protein (oppA) |
| 1B58 | 1.80 | 1RKM | 2.40 | KYK | 16 | Oligopeptide binding protein (oppA) |
| 1B5I | 1.90 | 1RKM | 2.40 | KNK | 16 | Oligopeptide binding protein (oppA) |
| 1B5J | 1.80 | 1RKM | 2.40 | KQK | 17 | Oligopeptide binding protein (oppA) |
| 1B9J | 1.80 | 1RKM | 2.40 | KLK | 16 | Oligopeptide binding protein (oppA) |
| 1JET | 1.20 | 1RKM | 2.40 | KAK | 14 | Oligopeptide binding protein (oppA) |
| 1JEU | 1.25 | 1RKM | 2.40 | KEK | 17 | Oligopeptide binding protein (oppA) |
| 1JEV | 1.30 | 1RKM | 2.40 | KWK | 16 | Oligopeptide binding protein (oppA) |
| 1QKA | 1.80 | 1RKM | 2.40 | KRK | 18 | Oligopeptide binding protein (oppA) |
| 1QKB | 1.80 | 1RKM | 2.40 | KVK | 15 | Oligopeptide binding protein (oppA) |
| 1S2K | 2.00 | 1S2B | 2.10 | AIH | 10 | SCP – B a member of the Eqolisin family of peptidases |
| **Tetra-peptide complexes** | | | | | | |
| 1TJ9 | 1.10 | 1FB2 | 1.95 | VARS | 14 | Daboia russelli pulchella phospholipase A2 |
| 1TK4 | 1.10 | 1FB2 | 1.95 | AIRS | 15 | Daboia russelli pulchella phospholipase A2 |
| 1OLC | 2.10 | 1RKM | 2.40 | KKKA | 20 | Oligopeptide binding protein (oppA) |
| 2DQK | 1.93 | 2PKC | 1.50 | VLLH | 15 | Proteinase K |
| 2FIB | 2.10 | 1FIB | 2.10 | GPRP | 11 | Recombinant human gamma-fibrinogen carboxyl terminal |
| 2NPH | 1.65 | 1G6L | 1.90 | AETF | 14 | HIV-1 protease |
| 1NRS | 2.40 | 1MKX | 2.20 | LDPR | 15 | Human alpha thrombin |
| **Penta-peptide complexes** | | | | | | |
| 1BHX | 2.30 | 1MKX | 2.20 | DFEEI | 22 | Human alpha thrombin |
| 1TJK | 1.25 | 1FB2 | 1.95 | FLSTK | 20 | Daboia russelli pulchella phospholipase A2 |
| 1TG4 | 1.70 | 1FB2 | 1.95 | FLAYK | 20 | Daboia russelli pulchella phospholipase A2 |
| 1JQ9 | 1.80 | 1FB2 | 1.95 | FLSYK | 21 | Daboia russelli pulchella phospholipase A2 |
| 1MF4 | 1.90 | 1SZ8 | 1.50 | VAFRS | 18 | Naja naja sagittifera phospholipase A2 |
| 2DUJ | 1.67 | 2PKC | 1.50 | LLFND | 20 | Proteinase K |
| 2GNS | 2.30 | 1FB2 | 1.95 | ALVYK | 19 | Daboia russelli pulchella phospholipase A2 |
| **Hexa-peptide complexes** | | | | | | |
| 1TP5 | 1.54 | 3I4W | 1.35 | KKETWV | 27 | PDZ3 domain of PSD-95 |
| 1OBY | 1.85 | 1NTE | 1.24 | TNEFYA | 22 | PDZ2 of syntenin |
| 2FOP | 2.1 | 1YZE | 2.00 | EKPSSS | 21 | Ubiquitin carboxyl-terminal hydrolase |
| 2FOO | 2.2 | 1YZE | 2.00 | EPGGSR | 19 | Ubiquitin carboxyl-terminal hydrolase |
| **Hepta-peptide complexes** | | | | | | |
| 1P7W | 1.02 | 2PKC | 1.50 | PAPFASA | 15 | Proteinase K |
| 1P7V | 1.08 | 2PKC | 1.50 | PAPFAAA | 14 | Proteinase K |
| 1U8I | 2.00 | 2F5A | 2.05 | ELDKWAN | 29 | HIV-1 cross neutralizing monoclonal antibody 2F5 |
| 1KY6 | 2.00 | 1B9K | 1.90 | FSDPWGG | 19 | Alpha-adaptin C |
| 2FOJ | 1.6 | 1YZE | 2.00 | GARAHSS | 22 | Ubiquitin carboxyl-terminal hydrolase |
| **Octa-peptide complex** | | | | | | |
| 2HD4 | 2.15 | 2PKC | 1.50 | GDEQGENK | 33 | Proteinase K |

**Table 1.** The 44 protein-peptide complexes used as test cases in this study, grouped by the length of the peptide ligand

## Results and discussions

We generated 1,500 docked structures for each test case of the benchmarking dataset. In this discussion, we specifically pick two structures from these. The first structure we identify for our analysis is what we call 'the best sampled structure'. For each test case, we consider the peptide structure in the respective *holo* protein-peptide complex as the native structure. The 1,500 peptide-protein complex structures generated using MOLS were each superimposed on the native structure without further changing the orientation. The RMSD was calculated on the peptide backbone alone. This procedure measures the differences not only in the structure of the peptide, but also in its pose with respect to the protein. We identify the best sampled structure as one in which the peptide has the lowest backbone RMSD with respect to the native structure.

| PDB ID | Peptide Sequence | Native energy | Lowest Energy[a] | | | Best Sampled[d] | | | Ranking[e] (%) |
|---|---|---|---|---|---|---|---|---|---|
| | | | RMSD[b] (Å) | Energy (no unit) | RDE[c] (%) | RMSD (Å) | Energy (no unit) | RDE (%) | |
| 8GCH | GAW | -870.81 | 1.69 | -949.43 | 53.11 | 1.62 | -859.77 | 46.78 | 4.20 |
| 1A30 | EDL | -886.99 | 5.83 | -792.56 | 75.09 | 1.10 | -713.90 | 57.56 | 11.50 |
| 1B32 | KMK | -1481.97 | 5.55 | -1032.87 | 74.00 | 1.72 | -896.85 | 60.14 | 35.70 |
| 1B05 | KCK | -1395.40 | 5.84 | -1029.37 | 72.72 | 2.09 | -924.41 | 60.37 | 12.20 |
| 1B3F | KHK | -1464.94 | 3.03 | -1077.60 | 64.00 | 1.79 | -968.99 | 58.01 | 12.60 |
| 1B3L | KGK | -1254.63 | 6.00 | -988.15 | 72.58 | 1.81 | -816.13 | 61.97 | 61.30 |
| 1B3G | KIK | -1394.79 | 3.66 | -1041.50 | 66.30 | 1.98 | -919.84 | 57.18 | 9.60 |
| 1B46 | KPK | -1428.67 | 3.39 | -944.46 | 62.89 | 2.05 | -661.75 | 62.33 | 69.40 |
| 1B4Z | KDK | -1447.42 | 6.00 | -1053.49 | 74.89 | 1.96 | -881.66 | 63.54 | 49.80 |
| 1B51 | KSK | -1423.22 | 6.53 | -1080.75 | 74.21 | 1.89 | -890.99 | 59.53 | 39.30 |
| 1B58 | KYK | -1516.95 | 5.50 | -1089.65 | 70.73 | 1.88 | -1014.91 | 59.88 | 9.07 |
| 1B5I | KNK | -1465.83 | 6.43 | -1076.80 | 74.70 | 2.06 | -850.59 | 62.69 | 71.70 |
| 1B5J | KQK | -1504.96 | 4.01 | -1066.40 | 70.61 | 1.97 | -520.93 | 64.44 | 97.90 |
| 1B9J | KLK | -1441.99 | 6.44 | -1047.52 | 75.09 | 1.95 | -928.64 | 62.27 | 21.30 |
| 1JET | KAK | -1323.79 | 4.70 | -1033.05 | 66.73 | 1.87 | -798.73 | 61.86 | 77.90 |
| 1JEU | KEK | -1444.22 | 4.68 | -1062.08 | 74.31 | 1.92 | -841.29 | 63.00 | 69.00 |
| 1JEV | KWK | -1454.60 | 6.43 | -1131.75 | 72.68 | 1.89 | -1003.52 | 60.73 | 15.90 |
| 1QKA | KRK | -1551.75 | 6.29 | -1062.55 | 72.74 | 2.00 | -813.83 | 68.90 | 73.80 |
| 1QKB | KVK | -1459.48 | 6.49 | -1071.12 | 73.81 | 1.80 | -880.55 | 59.31 | 52.00 |
| 1S2K | AIH | -600.46 | 5.80 | -715.40 | 72.11 | 1.50 | -608.98 | 61.31 | 15.70 |
| 1TK4 | AIRS | -774.76 | 5.78 | -682.09 | 68.73 | 2.21 | -543.39 | 63.39 | 2.80 |
| 1TJ9 | VARS | -571.31 | 6.10 | -658.96 | 72.60 | 1.46 | -463.23 | 56.60 | 55.90 |
| 2FIB | GPRP | -717.24 | 6.48 | -789.57 | 75.29 | 1.63 | -615.93 | 56.46 | 30.70 |
| 2DQK | VLLH | -171.34 | 5.26 | -957.89 | 67.21 | 1.99 | -796.92 | 61.78 | 49.70 |
| 2NPH | AETF | -827.63 | 5.52 | -906.61 | 71.65 | 1.54 | -839.45 | 51.93 | 1.30 |
| 1OLC | KKKA | -1560.03 | 2.74 | -1145.00 | 69.82 | 2.10 | -760.29 | 76.36 | 82.40 |
| 1NRS | LDPR | -822.17 | 4.92 | -489.16 | 71.52 | 1.63 | -455.34 | 57.42 | 0.73 |
| 2DUJ | LLFND | -1256.21 | 8.54 | -1058.33 | 73.76 | 3.21 | -1026.52 | 59.94 | 0.50 |
| 1MF4 | VAFRS | -745.03 | 5.68 | -1289.01 | 70.35 | 2.28 | -1110.16 | 66.52 | 12.73 |
| 1TJK | FLSTK | -700.96 | 8.23 | -447.40 | 79.02 | 2.71 | -224.57 | 65.98 | 49.53 |
| 1TG4 | FLAYK | 4109.31[f] | 4.67 | -473.24 | 70.19 | 2.51 | -206.20 | 61.28 | 49.33 |
| 1JQ9 | FLSYK | -928.87 | 1.92 | -567.13 | 62.21 | 1.64 | -399.28 | 62.75 | 2.47 |
| 2GNS | ALVYK | 35.04 | 9.58 | -867.40 | 76.57 | 1.93 | -202.37 | 60.00 | 31.93 |
| 1BHX | DFEEI | -659.31 | 7.98 | -440.13 | 78.97 | 2.01 | -230.46 | 59.05 | 27.40 |
| 1OBY | TNEFYA | -1412.04 | 8.69 | -878.12 | 78.48 | 5.23 | -475.11 | 73.09 | 94.30 |
| 1TP5 | KKETWV | -1032.59 | 2.81 | -821.69 | 64.36 | 1.95 | -755.57 | 60.50 | 0.40 |
| 2FOO | EPGGSR | -905.22 | 3.30 | -1258.99 | 64.21 | 1.80 | -1063.24 | 54.52 | 25.60 |
| 2FOP | EKPSSS | -943.37 | 3.81 | -1345.65 | 66.61 | 2.41 | -1212.60 | 62.88 | 10.87 |
| 1KY6 | FSDPWGG | -1037.78 | 7.08 | -1036.66 | 66.91 | 2.93 | -977.11 | 63.98 | 1.07 |
| 1U8I | ELDKWAN | -1045.50 | 9.83 | -1080.34 | 79.56 | 2.15 | -982.20 | 61.35 | 1.80 |
| 1P7V | PAPFAAA | 8705.57[g] | 7.15 | -902.98 | 73.05 | 2.55 | -738.16 | 57.56 | 10.40 |
| 2FOJ | GARAHSS | -1111.47 | 8.90 | -1756.19 | 74.41 | 2.78 | -1553.74 | 61.06 | 42.40 |
| 1P7W | PAPFASA | -441.29 | 4.97 | -851.07 | 71.58 | 2.65 | -781.21 | 57.94 | 2.40 |
| 2HD4 | GDEQGENK | -909.21 | 11.20 | -1179.33 | 83.03 | 3.19 | -815.53 | 66.60 | 79.73 |

[a] The peptide-protein complex with the lowest energy which is also the top-ranked structure.
[b] Root mean square deviation (RMSD) of the peptide docking pose (peptide conformation and peptide orientation) with respect to the native peptide docking pose.
[c] RDE – Relative displacement error - a metric used to rank different conformations and poses of a ligand docked to the receptor with respect to the known correct solution
[d] 'Best sampled' is the peptide docking pose with the lowest RMSD with respect to the native peptide docking pose.
[e] Rank of the best sampled solution among the energy-ranked solutions.
[f] High native energy due to high intra-peptide energy

[g] High native energy due to high protein-ligand intermolecular energy

**Table 2.** Summary of results obtained from iMOLSDOCK for the benchmark dataset of 44 test cases

The second structure we discuss from the 1,500 generated structures is the one with the lowest total energy, i.e. the top-ranked structure.

***Sampling and scoring efficiency in iMOLSDOCK***

Table 2 summarizes the backbone RMSD, Relative Displacement Error (RDE) and docking energy for the best sampled and the lowest energy structure obtained by iMOLSDOCK in all 44 test cases. Figure 2 shows the best sampled structures of all the test cases. (RDE is a metric used to rank different conformations and poses of a ligand docked to the receptor with respect to the known correct solution[43]). In 36 of the 44 cases, the RMSD of the best sampled structures are less than or equal to 2.50 Å. The ligand RMSD range is between 1.10 Å and 11.20 Å, in general being larger for the larger peptides. We note, once again, that the RMSD represents not only the difference in the conformation of the peptide ligand with respect to the native structure, but also the difference in the pose of the ligand in the receptor site.

An ideal scoring function must be able to top-score the best sampled structure. In the present case, in 27% (12/44) of the total test cases the best sampled structures were ranked within the top 10% energy ranked structures, and in an additional 18% (8/44) they were ranked between the top 10% and 20% of the energy ranks. In 2 cases, the top-scored structures were within 2.00 Å RMSD of the native structure. iMOLSDOCK did not find exact solution – the lowest RMSD structure having the lowest energy – in any of the 44 cases.

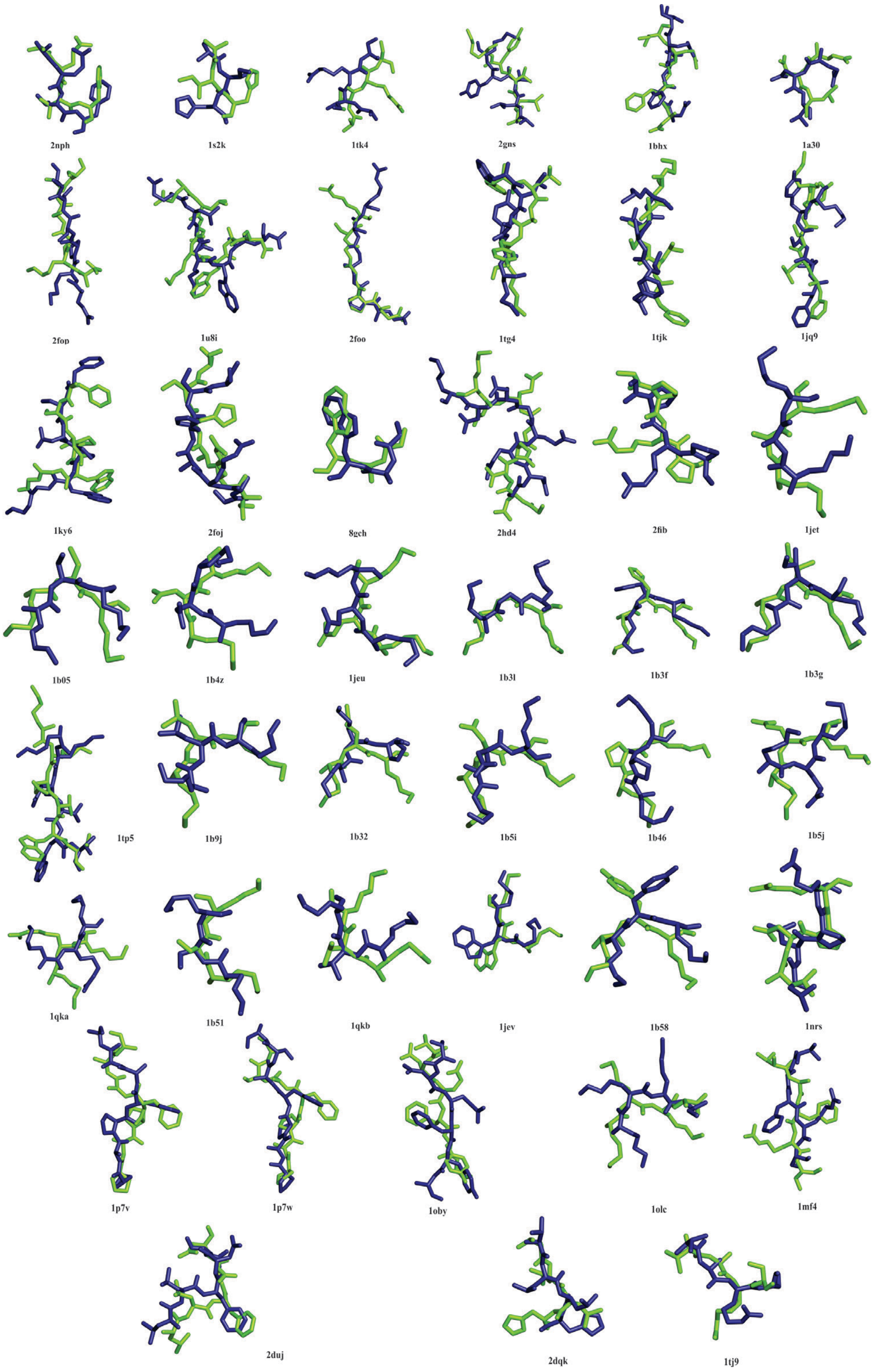

**Figure 2.** Best sampled structures obtained from iMOLSDOCK for the benchmark dataset of 44 test cases. For all the cases, the best sampled peptide structure (blue) is superimposed on the native peptide structure (green).

***Non-bonded interactions in iMOLSDOCK***

We have used HBPLUS[44] to find the non-bonded interactions between the protein and the ligand for iMOLSDOCK results. In Figure 3, we have compared the non-bonded interactions of the native complex and the top-ranked complex from iMOLSDOCK. In the latter set, 8 of the 44 cases have the number of hydrogen bonds greater than or equal to the number of hydrogen bonds in the crystal structure. In 24 of the 44 cases, iMOLSDOCK predicted more hydrophobic interactions in the top-ranked structure than in the native structure. Out of the 20 tripeptide cases, 17 are complexes of the sequence KxK (x stands for the variable residue) with the oligopeptide-binding protein (oppA). In all the 17 cases, the number of hydrogen bond interactions predicted by iMOLSDOCK in the top-ranked structure are less than the number of hydrogen bond interactions in the crystal structure. But in 9 of the 17 cases (1B05, 1B31,1B38, 1B4Z,1B51, 1B58, 1B5J, 1JET and 1QKB), and also in the remaining 3 tripeptide cases (8GCH, 1A30 and 1S2K) the hydrophobic interactions predicted in the top-ranked structure of iMOLSDOCK are greater than found in the crystal structure. The above results reveal that, in iMOLSDOCK, prediction of hydrophobic contacts is better than prediction of hydrogen bond interactions. The hydrogen bond prediction in iMOLSDOCK may probably be improved if we tune the weight of 'hydrogen bond interaction' term in the PLP term – the protein-ligand intermolecular interaction energy – of the iMOLSDOCK scoring function.

In 1OLC, iMOLSDOCK predicted 5 hydrogen bond interactions in the top-ranked structure which is very low compared to the 19 hydrogen bonds seen in the crystal structure. But the top-ranked structure (Energy = -1145.00) is energetically indistinguishable with the native structure (-1560.03). The number of hydrophobic interactions in the top-ranked structure and the native structure are almost same with 49 interactions and 48 interactions respectively. Therefore, the favourable energy in the top-ranked structure may be due to the 3 hydrogen bond interactions in the top-ranked structure (with protein residues TYR 109, GLU 32 and ASP 419) that are also seen in the native structure.

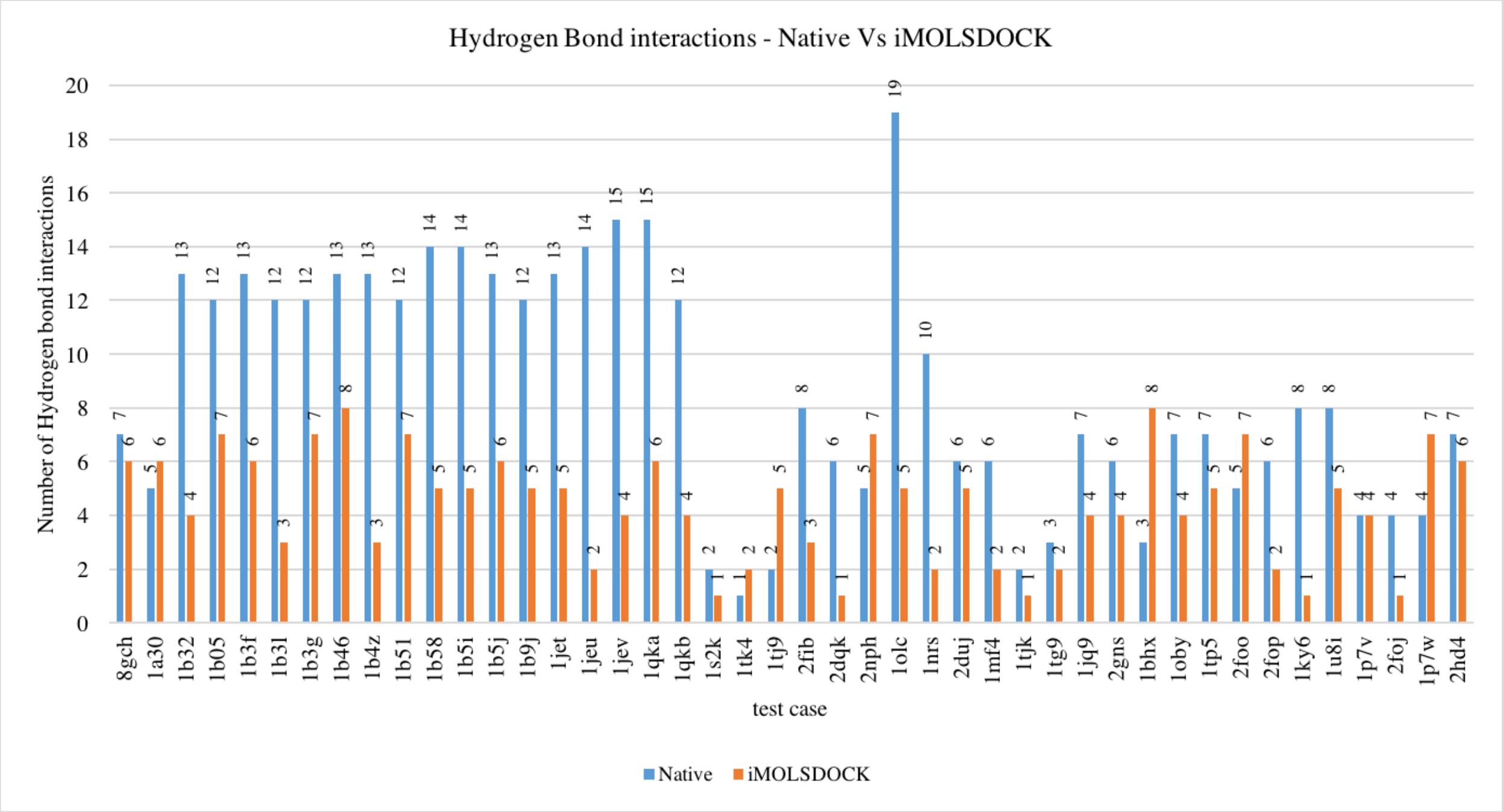

**Figure 3 (a).** Comparison of hydrogen bond interactions in the top-ranked iMOLSDOCK structure with the hydrogen bond interactions found in the native structure.

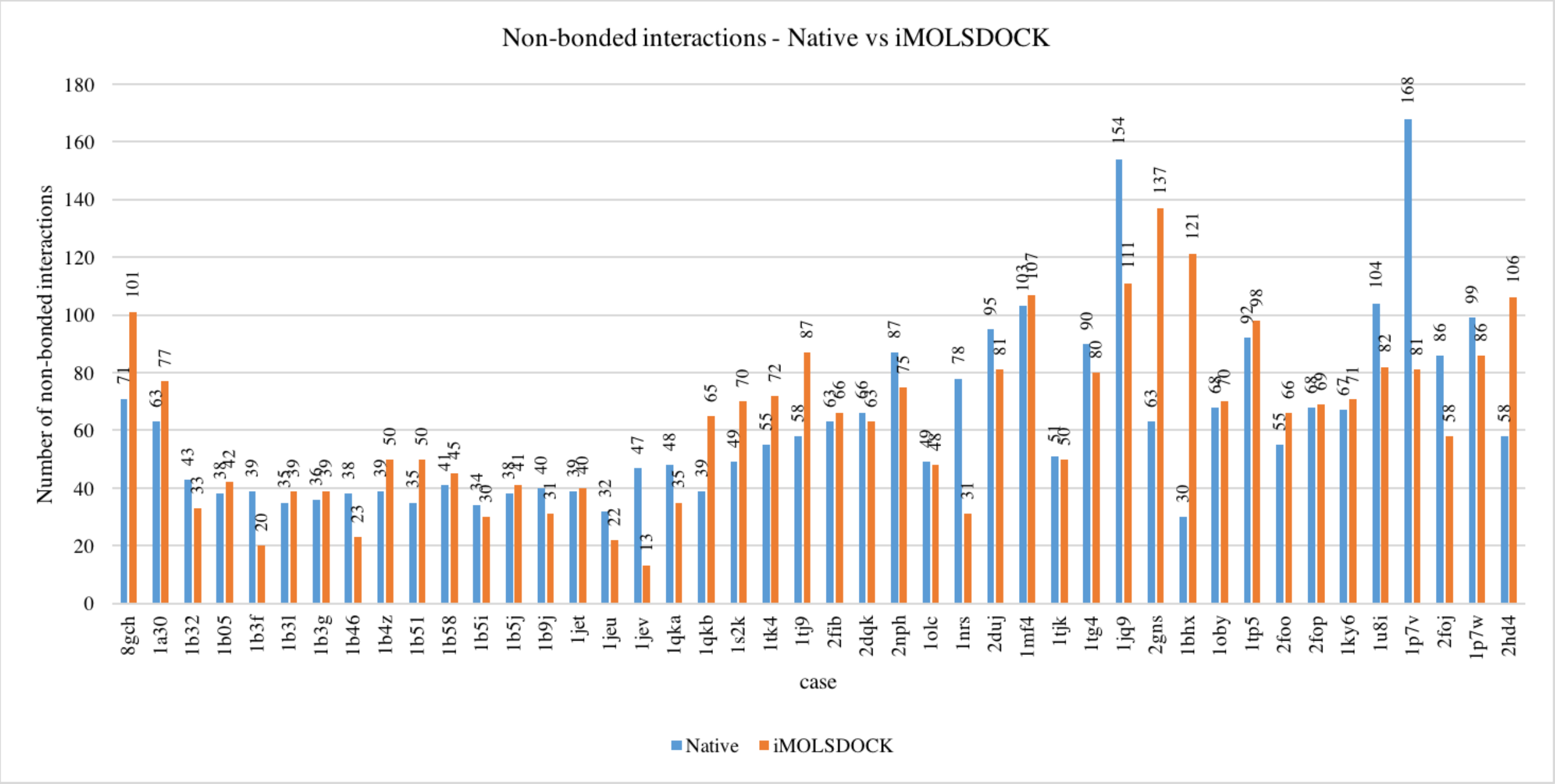

**Figure 3 (b).** Comparison of non-bonded interactions in the top-ranked iMOLSDOCK structure with the non-bonded interactions found in the native structure.

### *Alternate binding modes in iMOLSDOCK*

The method generates hundreds of low-energy possibilities and does not converge to just a single solution. Therefore alternate solutions that have a lower energy value than the native structure are often detected[45]. In some cases, the native, the best sampled and the lowest energy structures are energetically indistinguishable, i.e. they may be considered approximately iso-energetic[46]. Of the present results, in 10 test cases the lowest energy structure (i.e. the top ranked structure) shows an equally favourable alternate binding mode (Figure 4). These alternate binding modes are characterized by energy values lower than native energy, but large values of RMSD as compared to the native complex.

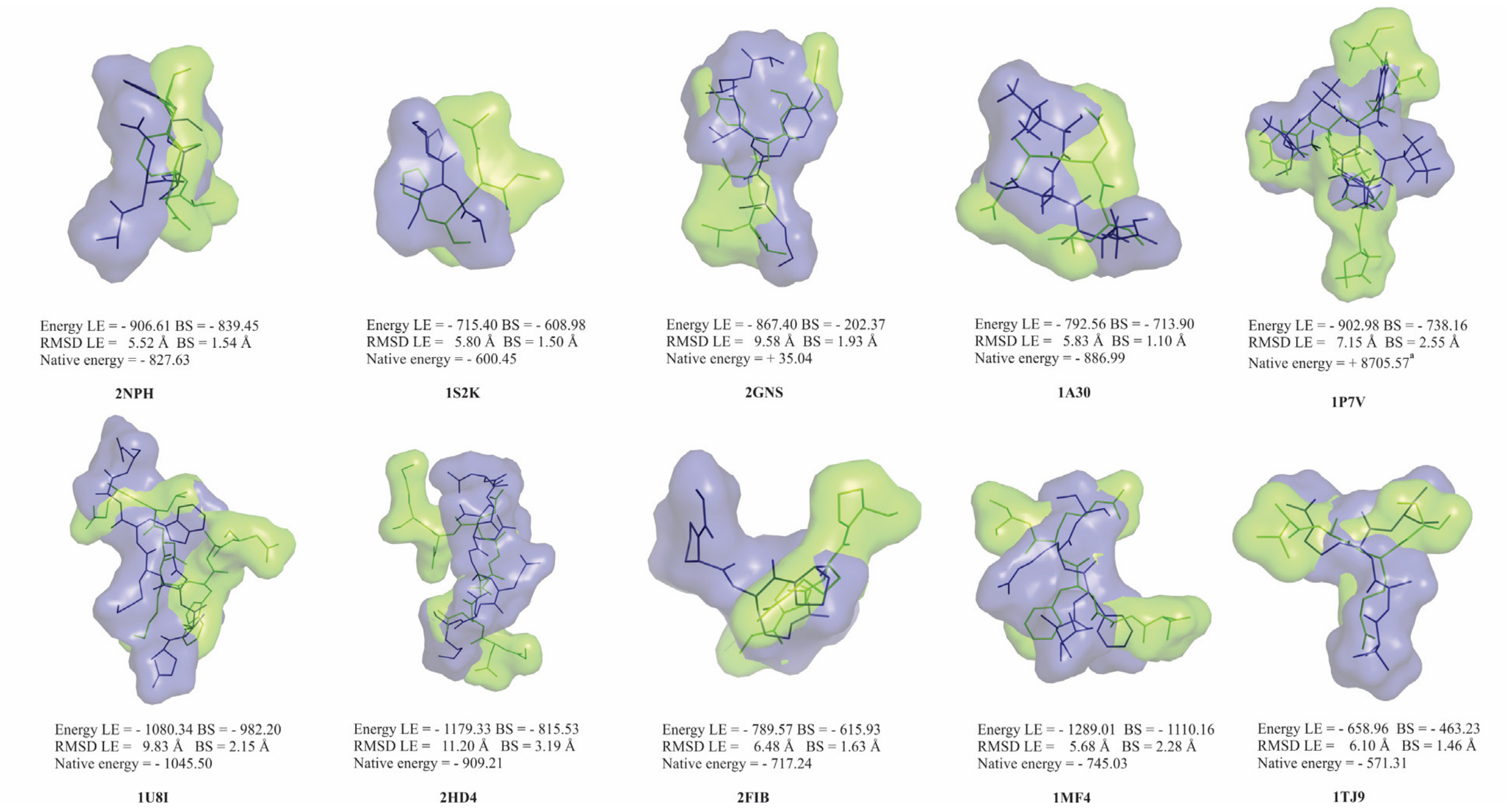

**Figure 4.** Alternate binding modes predicted by iMOLSDOCK top-ranked structures. The top-ranked structure of iMOLSDOCK is shown in blue and the native structure is shown in green. The native energy in 1P7V is high due to short contacts between the peptide and the protein.

Of particular note is the structure 2NPH. In this test case, the top-ranked structure has a larger number of hydrogen bond interactions than seen in the native complex. However, the RMSD of the top ranked structure with respect to the crystal structure is 5.52 Å. In 2NPH the top ranked structure predicted more hydrogen bond interactions (hydrogen bonds = 7) than the native structure (hydrogen bonds = 5). Figure 5 shows the hydrogen bond interactions of 2NPH common to the native structure and the top ranked structure.

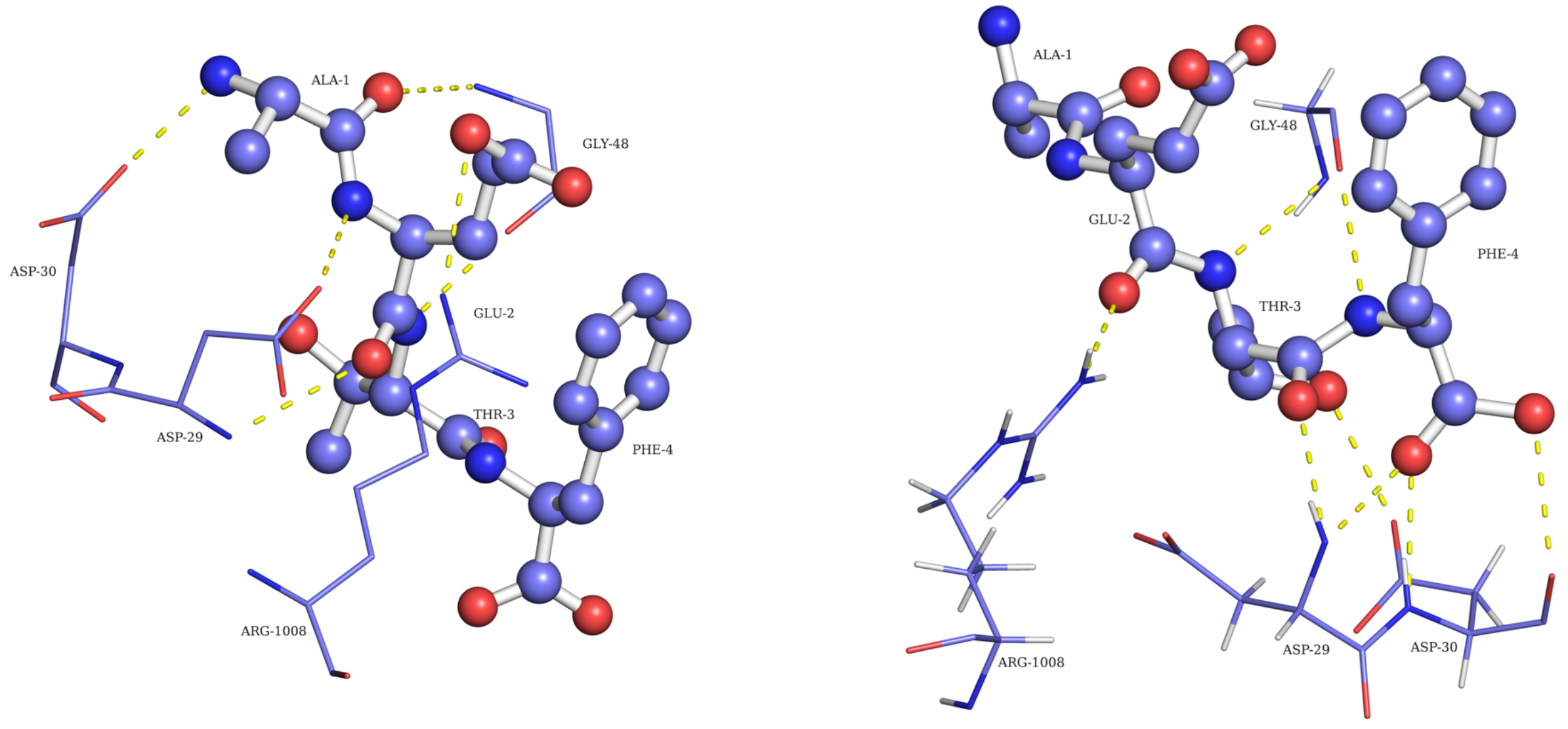

Energy = - 827.63 (no unit)

Binding mode in crystal structure

Energy = - 839.45 (no unit) RMSD = 5.52 Å

Alternate binding mode predicted by iMOLSDOCK

**Figure 5.** Alternate binding mode predicted by iMOLSDOCK for case 2NPH. Hydrogen bond interactions commonly found in the native and the top-ranked structure are shown in the above figure.

***Comparison of iMOLSDOCK with GOLD and AutoDock Vina***

We compared the results of iMOLSDOCK with GOLD, a commercial docking tool[17] and with AutoDock Vina, a free and commonly used docking tool[16] for the 44 cases.

| case | Vina | | | | GOLD | | | | Ranking[4] % | | |
|---|---|---|---|---|---|---|---|---|---|---|---|
| | Energy[1] (kcal/mol) | | RMSD[2] (Å) | | Score[3] (no unit) | | RMSD[2] (Å) | | | | |
| | Native | 1st rank | 1st rank | Best sampled | Native | 1st rank | 1st rank | Best sampled | Vina | GOLD | iMOLS-DOCK |
| 8GCH | - 2.58 | - 6.7 | 3.93 | 3.93 | + 45.29 | + 58.73 | 6.31 | 5.09 | EXACT | 52.6 | 4.2 |
| 1A30 | - 4.24 | - 7.0 | 5.47 | 3.04 | + 29.70 | + 48.86 | 5.49 | 2.43 | 100.0 | 53.0 | 11.5 |
| 1B32 | - 4.42 | - 6.3 | 7.74 | 4.14 | + 61.07 | + 70.77 | 5.55 | 2.75 | 88.8 | 40.0 | 35.7 |
| 1B05 | - 3.68 | - 6.4 | 6.99 | 5.80 | + 57.26 | + 61.13 | 5.41 | 3.00 | 44.4 | 94.0 | 12.2 |
| 1B3F | - 4.39 | - 7.3 | 6.82 | 6.03 | + 62.01 | + 69.53 | 5.19 | 2.10 | 55.5 | 62.6 | 12.6 |
| 1B3L | -3.52 | -7.0 | 6.95 | 4.85 | + 50.05 | + 56.81 | 6.23 | 4.32 | 22.2 | 96.0 | 61.3 |
| 1B3G | - 4.50 | - 6.5 | 8.61 | 3.97 | + 57.92 | + 62.37 | 5.44 | 2.56 | 66.6 | 69.3 | 9.6 |
| 1B46 | - 5.10 | - 6.7 | 6.72 | 4.04 | + 59.32 | + 52.83 | 4.57 | 4.57 | 33.3 | EXACT | 69.4 |
| 1B4Z | - 4.51 | - 6.9 | 6.00 | 5.24 | + 56.66 | + 53.42 | 5.33 | 2.66 | 44.4 | 15.3 | 49.8 |
| 1B51 | - 4.28 | - 6.8 | 6.48 | 4.96 | + 54.09 | + 47.89 | 5.38 | 2.42 | 33.3 | 90.6 | 39.2 |
| 1B58 | - 5.96 | - 7.3 | 4.78 | 3.74 | + 63.55 | + 56.54 | 6.38 | 2.87 | 66.6 | 76.0 | 9.1 |
| 1B5I | - 4.63 | - 7.3 | 6.90 | 5.19 | + 56.57 | + 47.49 | 5.82 | 2.83 | 44.4 | 66.6 | 71.7 |
| 1B5J | - 4.81 | - 7.1 | 5.62 | 4.03 | + 62.33 | + 53.28 | 6.23 | 2.79 | 33.3 | 89.3 | 97.9 |
| 1B9J | - 4.66 | - 6.8 | 7.55 | 4.03 | + 56.03 | + 50.32 | 6.84 | 2.48 | 88.8 | 29.3 | 21.2 |
| 1JET | - 4.79 | - 7.1 | 7.32 | 4.97 | + 51.74 | + 55.72 | 4.36 | 3.02 | 44.0 | 19.3 | 77.8 |
| 1JEU | - 4.98 | - 6.5 | 6.92 | 5.81 | + 57.08 | + 62.47 | 6.32 | 2.78 | 77.7 | 2.6 | 69.0 |
| 1JEV | - 5.66 | - 7.6 | 7.70 | 4.17 | + 68.66 | + 70.28 | 6.05 | 2.51 | 88.8 | 32.0 | 15.9 |
| 1QKA | - 5.06 | - 6.9 | 7.08 | 4.04 | + 72.13 | + 76.10 | 6.95 | 2.58 | 77.7 | 26.6 | 73.8 |
| 1QKB | - 4.28 | - 6.5 | 7.41 | 3.94 | + 22.11 | + 61.21 | 6.17 | 2.10 | 22.2 | 3.3 | 52.0 |
| 1S2K | - 4.12 | - 6.7 | 1.21 | 1.21 | + 29.52 | + 54.54 | 4.45 | 4.45 | EXACT | EXACT | 15.7 |
| 1TK4 | - 2.17 | - 7.4 | 6.92 | 6.55 | + 36.56 | + 61.71 | 7.47 | 5.06 | 88.8 | 51.3 | 2.8 |
| 1TJ9 | - 0.31 | - 8.1 | 10.27 | 9.68 | + 24.94 | + 31.81 | 10.94 | 6.92 | 66.6 | 66.0 | 55.9 |
| 2FIB | - 4.78 | - 8.6 | 5.95 | 3.92 | + 43.33 | + 53.01 | 7.33 | 5.52 | 33.3 | 42.6 | 30.7 |
| 2DQK | - 1.46 | - 6.7 | 8.32 | 3.21 | + 35.48 | + 71.73 | 7.40 | 3.10 | 22.2 | 89.3 | 49.7 |
| 2NPH | - 4.13 | - 7.9 | 3.46 | 1.44 | + 35.96 | + 56.55 | 4.52 | 2.01 | 33.3 | 23.3 | 1.3 |
| 1OLC | - 4.87 | - 6.6 | 8.58 | 3.19 | + 75.87 | + 69.77 | 7.43 | 5.16 | 100.0 | 10.6 | 82.4 |
| 1NRS | - 3.62 | - 5.8 | 6.95 | 6.95 | + 56.11 | + 79.07 | 8.36 | 3.45 | EXACT | 24.6 | 0.7 |
| 2DUJ | - 1.72 | - 7.1 | 5.97 | 4.38 | + 26.04 | + 70.92 | 7.32 | 5.03 | 22.2 | 83.3 | 0.5 |
| 1MF4 | + 1.34 | - 8.0 | 8.53 | 6.40 | - 22.43 | + 73.59 | 8.09 | 4.19 | 100 | 83.3 | 12.73 |
| 1TJK | - 0.72 | - 7.7 | 7.60 | 6.76 | + 7.28 | + 82.13 | 7.05 | 4.81 | 88.8 | 16.66 | 49.5 |
| 1TG4 | + 0.64 | - 7.4 | 9.19 | 5.83 | - 157.90 | + 82.33 | 5.96 | 3.78 | 77.7 | 12.66 | 49.3 |
| 1JQ9 | + 0.85 | - 7.4 | 9.95 | 6.70 | - 110.46 | + 84.10 | 9.13 | 2.64 | 66.6 | 60.0 | 2.47 |
| 2GNS | - 4.42 | - 9.4 | 10.83 | 10.06 | + 45.89 | + 88.67 | 10.94 | 3.51 | 44.4 | 10.0 | 31.9 |
| 1BHX | - 3.57 | - 5.3 | 4.77 | 4.62 | + 30.59 | + 71.92 | 8.64 | 5.65 | 66.6 | 14.0 | 27.4 |
| 1OBY | - 5.25 | - 6.3 | 9.10 | 8.39 | - 2700.65[c] | + 73.44 | 11.19 | 7.86 | 55.5 | 34.6 | 94.2 |
| 1TP5 | - 3.37 | - 6.0 | 11.13 | 10.17 | - 25.19 | + 70.39 | 7.60 | 5.70 | 44.4 | 2.6 | 0.4 |
| 2FOO | - 2.66 | - 5.7 | 7.00 | 5.58 | - 4131.54[c] | + 81.78 | 12.70 | 2.74 | 55.5 | 32.7 | 25.6 |
| 2FOP | - 3.03 | - 5.5 | 10.22 | 8.39 | - 9624.5[c] | + 70.61 | 8.09 | 5.60 | 88.8 | 92.0 | 10.8 |
| 1KY6 | - 4.72 | - 7.8 | 12.32 | 8.38 | - 1332.56[c] | + 69.93 | 7.92 | 4.33 | 33.3 | 17.3 | 1.07 |
| 1U8I | - 3.52 | - 6.1 | 7.42 | 6.77 | + 43.39 | + 74.52 | 7.16 | 4.14 | 22.2 | 24.0 | 1.8 |
| 1P7V | + 598.80 | - 8.2 | 8.82 | 5.17 | - 6042.07[c] | + 75.49 | 8.03 | 4.11 | 77.7 | 72.0 | 10.4 |
| 2FOJ | - 3.44 | - 4.1 | 9.36 | 6.20 | - 1327.70[c] | + 70.12 | 10.78 | 2.96 | 22.2 | 47.3 | 42.4 |
| 1P7W | - 1.52 | - 8.0 | 6.59 | 5.17 | + 42.19 | + 68.65 | 8.86 | 5.47 | 77.7 | 100.0 | 2.4 |
| 2HD4 | - 2.42 | - 5.9 | 4.74 | 4.05 | - 1351.01[c] | + 76.03 | 11.13 | 4.24 | 88.8 | 24.6 | 79.7 |

[c] The very high native energy is due to high van der Waals energy in the scoring function.
[1] In Vina, the more negative the energy, the better is the solution.
[2] Root mean square deviation (RMSD) of the peptide docking pose (peptide conformation and peptide orientation) with respect to the native peptide docking pose.
[3] In GOLD, the more positive the score the better is the solution.
[4] Rank of the best sampled solution among the energy-ranked solutions.

**Table 3.** Summary of results comparison between iMOLSDOCK, GOLD and AutoDock Vina

GOLD is a GA-based docking tool. The parameters of the fitness function were fixed to their default values. A total of 150 GA runs were performed for each test case. We set the docking speed to '*slow (most accurate)*' which equated to 100,000 operations. The rest of the options were set to 'automatic'.

AutoDock Vina is an open-source program for molecular docking[16]. It uses Iterated Local Search global optimizer algorithm where a succession of steps are taken, each consisting of a mutation and a local optimization. Each step is accepted or rejected according to the Metropolis criterion. The grid size for the search space was set to the optimum size mentioned by the authors[16], to accommodate the extended conformation of the peptide ligand inside the grid box. The centre of the grid box was set to the centre of the peptide. All other settings (exhaustiveness, number of modes, etc.) have been set to the default values. The input peptide structure for both GOLD and Vina was the extended structure built using the MOLS 2.0 - peptide modeling tool[30]. In iMOLSDOCK, the amino acid sequence of the peptide ligand is given as input; the extended structure of the peptide is automatically built using the *BUILDER* routine of the iMOLSDOCK algorithm. Table 3 summarises the backbone RMSD of the top-ranked structures, RMSD of the best sampled structures and energy of the top-ranked structures of GOLD and Vina. Table 3 also shows the ranking efficiency of Vina, GOLD and iMOLSDOCK.

Table 4 shows a comparison of the results from iMOLSDOCK, GOLD and Vina. The number of cases in which the prediction with lowest RMSD with respect to the native complex (i.e. the best sampled structure has the RMSD value less than 2.50 Å) is the greatest in the case of iMOLSDOCK. Clearly this program, iMOLSDOCK, is able to arrive at docked structures that are close to the crystal structure in a larger number of cases than the other two algorithms (GOLD and Vina). Likewise, iMOLSDOCK generates a larger number of predictions in which the structure with the lowest RMSD also falls in the set with the lowest energy. However, GOLD and AutoDock Vina are able to identify exact solutions – the lowest RMSD structure having the lowest energy – in two and three cases respectively, while iMOLSDOCK makes no such prediction.

| Parameter | iMOLSDOCK | GOLD | Vina |
|---|---|---|---|
| Best sampled solutions with RMSD ≤ 2.50 Å | 37 | 6 | 2 |
| Best sampled structures within top 10% energy ranking | 12 | 5 | NIL |
| Lowest RMSD among the top-ranked solution of all the cases | 1.69 Å | 1.21 Å | 4.36 Å |
| Highest RMSD among the top-ranked solution of all the cases | 11.20 Å | 12.32 Å | 12.70 Å |
| Average RMSD among the top-ranked solutions | 5.8 Å | 7.32 Å | 7.24 Å |
| Exact solutions[a] | NIL | 2 | 3 |
| Sampling range[b] - RMSD | 1.10 Å – 11.20 Å | 2.01 Å – 12.70 Å | 1.21 Å – 12.32 Å |

[a] Solutions for which the best sampled is the same as the top-ranked structure
[b] Sampling range taken from all the best sampled and top-ranked structures of all the cases

**Table 4.** Comparison between iMOLSDOCK, GOLD and Vina for the 44 test cases

In Vina, the best sampled solutions of all cases (except three), are at least 3.00 Å RMSD away from the crystal structure. In GOLD, for 20 cases the best sampled ligand is within 3.00 Å RMSD from the crystal structure. In iMOLSDOCK, except for the three cases 2DUJ, 1OBY and 2HD4, the remaining 41 cases have the best sampled ligand within 3.00 Å RMSD from the crystal structure. The main reason behind the success of iMOLSDOCK sampling and ranking over Vina may be due to systematic search technique approach and the large number of structures generated (1,500 structures) for each test case. In Vina we used the default number of binding modes (9 binding modes) and *exhaustiveness*. As part of the testing, for all the cases we tried increasing the number of binding modes (*num_modes*) and the *exhaustiveness* but the results were similar to that obtained with the default *num_modes* and *exhaustiveness*.

***Computation time***

Table 5 shows a comparison of the CPU time required by AutoDock Vina (9 binding modes), GOLD (150 runs) and iMOLSDOCK (1,500 runs) to find the solution. Also indicated is the total number of variable torsion angles (backbone and side-chain torsion angles of the peptide ligand + side chain torsion angles of the receptor protein) considered. We fixed 10 flexible residues in the receptor protein for both Vina and GOLD. For all the test cases in iMOLSDOCK, we enabled automatic selection of flexible residues. That is, the program automatically selected residues that are within 4.0 Å from each atom of the peptide ligand as the flexible residues. On an average up to 40 flexible residues were treated as flexible.

| | **Vina** | **GOLD** | **iMOLSDOCK** |
|---|---|---|---|
| **Search technique** | MC | GA | Systematic search |
| **Total runs** | 9 | 150 | 1500 |
| **Flexible residues** | 10 | 10 | 40 |
| **Protein torsion angles (average)** | 18 | 22 | 65 |
| **Total number of parameters (average)** | 37 | 37 | 93 |
| **Average time for total runs** | 0.6 h | 3.2 h | 4797.0 h |
| **Average time per structure** | 4.0 m | 2.0 m | 3.0 h |

MC – Monte Carlo, GA – Genetic Algorithm

**Table 5.** CPU time comparison between iMOLSDOCK, GOLD and AutoDock Vina

The average CPU time required for one run is 4 m for AutoDock Vina, 2 m for GOLD and 3 h for iMOLSDOCK. The CPU time of iMOLSDOCK is a function of number of peptide torsion angles, number of protein torsion angles and number of protein atoms in the docking space.

## Conclusions

We have enhanced the capability of the iMOLSDOCK docking algorithm by incorporating 'induced-fit' side-chain receptor flexibility. Test runs using 44 peptide-protein complexes from the PDB show that the program performs well. Comparison using the same test cases with GOLD and AutoDock Vina, shows that iMOLSDOCK outperforms these two commonly used programs, though it consumes an order of magnitude more computer time. It may be noted however, that not only is the search space vastly greater in iMOLSDOCK than in the other two, incorporating, as it does, a larger number of flexible residues, but also the search is systematic and not stochastic. An added advantage of the iMOLSDOCK algorithm, as pointed out earlier[6,7] is its ability to identify alternate binding modes. This paper reports iMOLSDOCK as demonstrated and tested for peptide ligands. However, it is straightforward to extend the algorithm it to other drug-like ligands.

## CONFLICT OF INTEREST

The authors declare no conflict of interest.

## ACKNOWLEDGMENT

We thank the Department of Science and Technology, Government of India, for financial support. We also thank the University Grants Commission for support under the CAS program.